\documentclass{PoS}
\usepackage{amsmath,amssymb,amsfonts,macros_static,amscd}

\title{Radiative improvement of spin and Darwin terms in the NRQCD action}
\ShortTitle{Radiative Improvement of NRQCD}

\author{
T.C.~Hammant$\,^{a}$, A.G.~Hart$\,^{b}$, \speaker{G.M.~von~Hippel}$\,^{c}$, R.R.~Horgan$\,^{a}$, C.J.~Monahan$\,^{d}$\\
$^a$~Department~of~Applied~Mathematics and Theoretical~Physics, University~of~Cambridge, Cambridge~CB3~0WA, UK\\
$^b$~Cray Exascale Research Initiative Europe, JCMB, King's~Buildings,
Edinburgh~EH9~3JZ, UK\\
$^c$~Institut~f\"ur~Kernphysik, Johannes~Gutenberg-Universit\"at~Mainz, 55099~Mainz, Germany\\
$^d$~Department of Physics, College of William and Mary, VA 23187-8795, USA\\
\email{hippel@kph.uni-mainz.de}}

\abstract{We present updated results for the radiative improvement of the
          $\sigma\cdot B$ term and the spin-dependent four-fermion terms in the
          lattice NRQCD action, and first results for the radiative corrections
          to the NRQCD Darwin term and spin-independent four-fermion terms.
          The spin-dependent terms have significant impact on getting the
          correct hyperfine splitting for both bottomonium and heavy-light
          mesons, while the spin-independent terms suffer from a conspiracy
          between lattice artifacts and severe IR divergences that complicates
          their evaluation.
         }

\FullConference{The 30th International Symposium on Lattice Field Theory - Lattice 2012,\\
		June 24-29, 2012\\
		Cairns, Australia}

\begin{document}


\def\NR{\textrm{NR}}
\def\be{\begin{equation}}
\def\ee{\end{equation}}
\def\bea{\begin{eqnarray}}
\def\eea{\end{eqnarray}}
\def\rme{\mathrm{e}}
\def\Psibar{\overline{\Psi}}
\def\sla#1{\not\!\!#1}
\def\calL{\mathcal{L}}
\def\calO{\mathcal{O}}

\section{Motivation}

Non-Relativistic QCD (NRQCD)
\cite{Lepage:1992tx}
has been used successfully to describe both quarkonia and $B_q$ mesons.
However, until recently, only the tree-level NRQCD action has been used.
What is now known is that the radiative improvement of the coefficient
of the $\sigma\cdot B$ operator in the NRQCD action and the inclusion of
spin-dependent four-fermion operators into the action have a significant effect
on the value of the bottomonium hyperfine splitting
\cite{Hammant:2011bt}.
Here we present an update on the improvement of the $\sigma\cdot B$ and
four-fermion operators, as well as first results for the radiative improvement
of the Darwin term in the NRQCD action.

\section{Matching NRQCD to QCD in Background Field Gauge}

The NRQCD action used by the HPQCD collaboration is
\[
S = \sum_{\vec{x},\tau} \psi^\dag(\vec{x},\tau)\left[\psi(\vec{x},\tau)-K(\tau)\psi(\vec{x},\tau)\right]
\]
with the kernel
\[
K(\tau) = \left(1-\frac{\delta H|_\tau}{2}\right)\left(1-\frac{H_0|_\tau}{2n}\right)^nU_4^\dag(\tau-1)\left(1-\frac{H_0|_{\tau-1}}{2n}\right)^2\left(1-\frac{\delta H|_{\tau-1}}{2}\right)\,,
\]
where
\[
H_0 = \frac{\Delta^{(2)}}{2M_0}\,,\quad
\delta H = -c_1\frac{(\Delta^{(2)})^2}{8M_0^3} + c_2\frac{ig}{8M_0^2} \left(\vec{\Delta}^\pm\cdot \vec{E} - \vec{E}\cdot\vec{\Delta}^\pm\right)
\]\[
-c_3\frac{g}{8M_0^2}\vec{\sigma}\cdot\left(\vec{\Delta}^\pm\times \vec{E} - \vec{E}\times\vec{\Delta}^\pm\right) -c_4\frac{g}{2M_0}\vec{\sigma}\cdot\vec{B}
 +c_5a^2\frac{\Delta^{(4)}}{24M_0} +c_6a\frac{(\Delta^{(2)})^2}{16nM_0^2}
\]
and $n\ge 3/(M_0a)$ is a stability parameter used to avoid the well-known
instability of the Euclidean Schr\"odinger (i.e. diffusion) equation.
The normalization of the operators is chosen such that the tree-level matching
of NRQCD to QCD (which can be performed e.g.\ by a Foldy-Wouthuysen-Tani
transformation) gives $c_i=1+\calO(\alpha_s)$.

In order to match NRQCD to QCD at the one-loop level, we have to determine
the radiative corrections to $c_i$ by demanding that some suitably chosen
set of renormalized S-matrix elements agree to one-loop accuracy when
calculated in QCD and NRQCD. For the quark bilinear terms in $\delta H$,
we can use quark scattering off a background field for this purpose.
Hence we are led to demand that QCD and NRQCD give the same effective
potential after non-relativistic reduction, i.e.\ that the following
diagram commutes:
\be\begin{CD}
\textrm{QCD}     @>\textrm{NR reduction}>\textrm{using $c_i$}>  {\textrm{tree-level}\atop\textrm{NRQCD}} \\
@V\textrm{1PI}VV                            @VV\textrm{1PI}V \\
\rm\Gamma        @>>\textrm{NR reduction}>  \rm\Gamma^{\rm NR} \\
\end{CD}\ee

\subsection{The Background Field Method}

The effective potential in the presence of a classical background field $\Phi$
is defined by
\[
\rme^{-\Gamma[\Phi]} = \int\limits_{\rm 1PI} D\phi \; \rme^{-S[\Phi+\phi]}\,,
\]
where the path integral over the quantum fluctuations $\phi$ only makes
sense in perturbation theory restricted to 1PI diagrams.

In a gauge theory, we can decompose the gauge potential into a background
and a quantum part as $A_\mu = B_\mu + gq_\mu$. BRST invariance of the classical
action $S$ guarantees that all $D\le 4$ operators appearing in the effective
action $\Gamma$ are gauge covariant, implying the renormalizability of the
theory. The $D>4$ operators that will appear in $\Gamma$ are, however, not
necessarily gauge covariant.

As an effective theory, NRQCD contains $D>4$ operators in the classical action
$S$, where gauge covariance can be imposed at tree level. At the loop level,
gauge covariance must also be retained to avoid serious complications (even
though the appearance of gauge-noncovariant operators would not render the
theory invalid in and of itself, keeping track of their gauge dependence would
be highly cumbersome). As has been shown long ago,
\cite{DeWitt:etc},
this can be achieved by using background field gauge (BFG).

Background field gauge is defined by the gauge fixing function
\[
f(A) = D^{B}_\mu q^\mu = (\partial_\mu + i B_\mu)q^\mu\,,
\]
and hence not only the propagator, but also the three- and four-gluon vertices
of the form $qqB$ and $qqBB$ are gauge-parameter dependent.

An important feature of BFG which we will exploit in the following is that it
leads to QED-like Ward identities and finite counterterms,
so that we can compute all diagrams numerically, and do not need to calculate
the gauge field renormalization.
This is practical in particular for checking the gauge-parameter independence
of our results (note that $c_4$ and $c_2$ should be gauge-parameter independent,
since they are directly related to physical mass splittings).

To employ the background field method on the lattice, the gauge link is
decomposed into the ordered product
\[
U_\mu(x) = \rme^{g_0q_\mu(x+\frac{1}{2}\hat{\mu})}\rme^{B_\mu(x+\frac{1}{2}\hat{\mu})}
\]
leading to a dependence of the Feynman rules on the number of background and
quantum fields ($qqq$, $Bqq$, $BBq$, etc.), as well as to the appearance of
different terms for different orderings ($Bqq$, $qBq$, $qqB$, etc.) contributing
to the same vertex.
Background field gauge (BFG) is defined by the gauge fixing function
\[
f(A) = D^{B}_\mu q_\mu(x) = \left[q_\mu(x)-\rme^{-B_\mu(x-\frac{\hat{\mu}}{2})}q_\mu(x-\hat{\mu})\rme^{B_\mu(x-\frac{\hat{\mu}}{2})}\right]
\]
which on the lattice affects all vertices with exactly two quantum gluons.

The automated derivation of Feynman rules with background fields, as well as
the use of BFG, has been implemented in the HiPPY/HPsrc packages for automated
lattice perturbation theory
\cite{Hippy:etc}.

Lattice gauge theories in BFG are renormalizable
\cite{Luscher:1995vs}.
Checking that the gauge dependences of our results match for individual terms,
and that their sum is non-trivially gauge independent,
gives us confidence in the correctness of the results.

\subsection{Matching the spin and Darwin terms}

\begin{figure}
\begin{center}
\includegraphics[width=0.49\textwidth,keepaspectratio=]{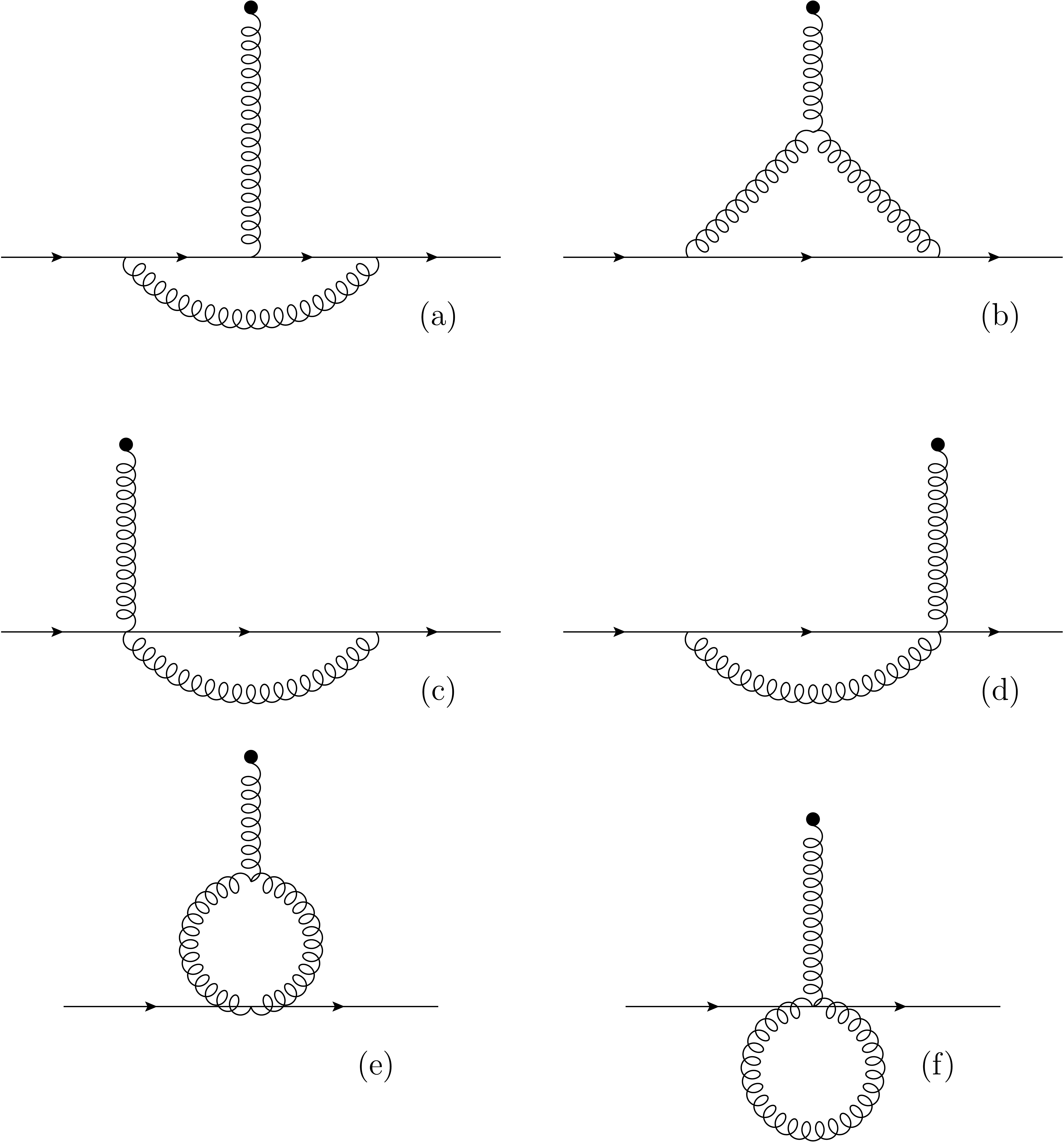}
\end{center}
\caption{The one-particle irreducible diagrams entering the matching of the
         $\sigma\cdot B$ and Darwin terms in NRQCD at the one-loop level.
         On the QCD side, only diagrams (a) and (b) need to be computed, while
         in NRQCD all diagrams contribute; note that diagrams (c)--(f) contain
         not only lattice artifacts, but also physical contributions from higher
         orders in $1/m$.}
\label{fig:bilindiag}
\end{figure}

The effective action for continuum QCD contains terms of the form
\[
\Gamma[\Psi,\Psibar,A] = \Psibar F_1(q^2)\sla{D}\Psi + \Psibar
\frac{F_2(q^2) \sigma^{\mu\nu}F_{\mu\nu}}{2M}\Psi + \ldots
\]
which after renormalization and non-relativistic reduction give
\[
(1+\underbrace{F_2(0)F_1(0)^{-1}}\limits_{=b_\sigma})\psi_R^\dag\frac{\vec{\sigma}\cdot\vec{B}}{2M_R}\psi_R
\]
for the $\sigma\cdot B$ term in $\Gamma^{\rm NR}$, and
\[
(1-\underbrace{8M_R^2F_1'(0)+2F_2(0)}\limits_{=b_D})\psi_R^\dag\left(-\frac{g q^2 A_0}{8M_R^2}\right)\psi_R\
\]
for the Darwin term.
A straightforward continuum calculation gives
\[
b_\sigma^{(1)} = \left(\frac{3}{2\pi}\log\frac{\mu}{M}+\frac{13}{6\pi}\right)\alpha  ~~~~~~~~~~
b_D^{(1)} = \left(-\frac{M^2}{\pi\mu^2}-\frac{7M}{4\mu}-\frac{1}{\pi}-\frac{50}{9\pi}\log\frac{\mu}{M}\right)\alpha
\]

On the other hand, the effective action for NRQCD contains the terms
\[
\Gamma[\psi,\psi^\dag,A] = c_4Z_\sigma^{\NR}\psi^\dag\frac{i\sigma\cdot(q\wedge A)}{2M}\psi -c_2Z_D^{\NR}\psi^\dag\frac{gq^2A_0}{8M^2}\psi+\ldots
\]
which after renormalization give
\[
\Gamma^{\rm NR}[\psi_R,\psi_R^\dag,A] =
  c_4Z_\sigma^{\NR}Z_2^{\NR}Z_m^{\NR}\psi_R^\dag\frac{i\sigma\cdot(q\wedge A)}{2M_R}\psi_R  -c_2Z_D^{\NR}Z_2^{\NR}(Z_m^{\NR})^2\psi_R^\dag\frac{gq^2A_0}{8M_R^2}\psi_R+\ldots
\]
Equating the corresponding terms in $\Gamma^{\rm NR}$ gives the matching
conditions
\bea
c_4 Z_\sigma^{\NR} Z_2^{\NR} Z_m^{\NR}     &=& 1 + b_\sigma \nonumber\\
c_2 Z_D^{\NR}      Z_2^{\NR} (Z_m^{\NR})^2 &=& 1 + b_D \nonumber
\eea
which yield $c_i^{(0)} = 1$ at tree-level, and at one-loop order (with $Z=1+\delta Z$)
\begin{eqnarray*}
c_2^{(1)} &=& b_D^{(1)}-\delta Z_D^{\NR,(1)}-\delta Z_2^{\NR,(1)}-2\delta Z_m^{\NR,(1)} \\
c_4^{(1)} &=& b_\sigma^{(1)}-\delta Z_\sigma^{\NR,(1)}-\delta Z_2^{\NR,(1)}-\delta Z_m^{\NR,(1)}
\end{eqnarray*}
In lattice NRQCD, we also need to take into account contributions from
mean-field improvement $U\mapsto U/u_0$ in $\delta Z_{\sigma,D,m}^{\NR,(1)}$
besides the diagrammatic contributions.

\subsection{Four-Fermion Operators}

\begin{figure}
\begin{center}
\includegraphics[width=0.5\textwidth,keepaspectratio=]{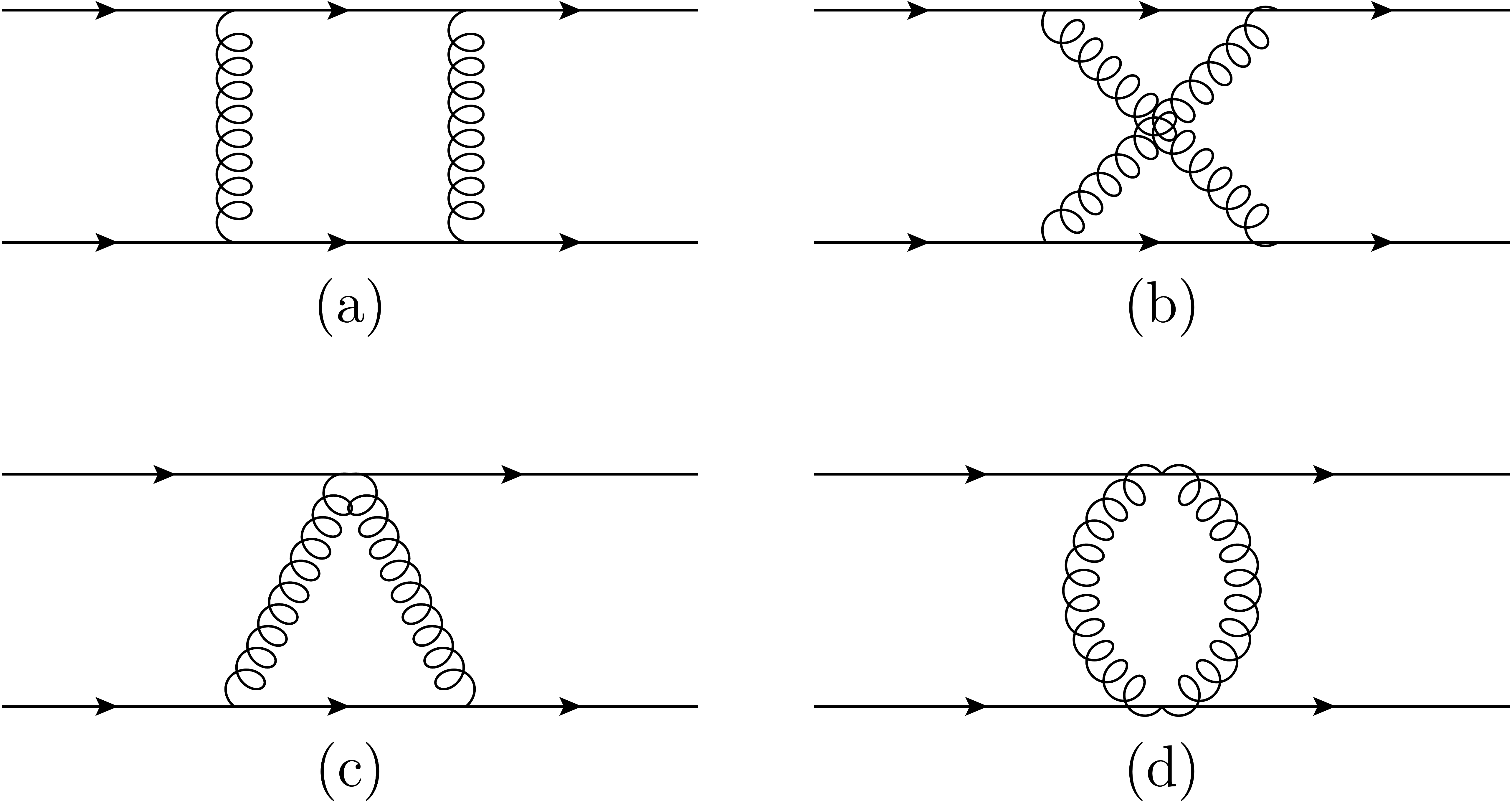}
\end{center}
\caption{The one-particle irreducible diagrams entering the matching of the
         four-fermion interactions in NRQCD. In QCD, only the box and cross-box
         diagrams contribute, while in NRQCD all diagrams must be computed.}
\label{fig:4fdiag}
\end{figure}

Beyond tree level, the NRQCD action also contains four-fermion terms
\begin{eqnarray*}
\calL_{4f} &=& d_1\frac{\alpha_s}{M^2}(\psi^\dag\chi^*)(\chi^t\psi)
            +d_2\frac{\alpha_s}{M^2}(\psi^\dag\sigma\chi^*)(\chi^t\sigma\psi)\\
           &&+d_3\frac{\alpha_s}{M^2}(\psi^\dag t^a\chi^*)(\chi^t t^a\psi)
            +d_4\frac{\alpha_s}{M^2}(\psi^\dag\sigma t^a\chi^*)(\chi^t\sigma t^a\psi)
\end{eqnarray*}
which can be rearranged by a Fierz transformations into
\begin{eqnarray*}
\calL_{4f} &=& a_1\frac{g^2}{M^2}(\chi^\dag\chi)(\psi^\dag\psi) 
            +a_8\frac{g^2}{M^2}(\chi^\dag t_a^t\chi)(\psi^\dag t_a\psi)\\
            &&+b_1\frac{g^2}{M^2}(\chi^\dag\sigma^*\chi)(\psi^\dag\sigma\psi)
            +b_8\frac{g^2}{M^2}(\chi^\dag\sigma^*t_a^t\chi)(\psi^\dag\sigma t_a\psi)
\end{eqnarray*}
where the coefficients $d_i$ are linear combinations of the coefficients
$a_i$, $b_i$, with the latter directly computable from the box diagrams
of fig.~\ref{fig:4fdiag}.

In addition to the box diagrams, there are additional contributions to
$d_i$ from $Q\overline{Q}$ annihilation, which is possible in QCD, but not
in NRQCD. For our purposes, the relevant contribution is
\[
d_1^{ann} = -\frac{2\alpha_s}{9M^2}(2-2\log 2).
\]

\section{Divergences and artifacts}

The calculated quantities in both QCD and NRQCD are IR divergent.
We can use a gluon mass $\mu$ to regulate these divergences,
since in the absence of ghost loops, all gluon lines are attached to a
conserved current removing the unphysical longitudinal mode contributions.
The NRQCD contribution then contains an IR logarithm $\log(\mu a)$, which
combines with the IR logarithm $\log(\mu/M)$ from the QCD side to give
the expected logarithmic $\log(M a)$-dependence for the matching coefficient.

On the other hand, any power divergences must match between QCD and NRQCD.
The spin-independent part contains power IR divergences up to $M^3/\mu^3$,
whereas the spin-dependent part is much more mildly divergent, with IR
divergences only up to $M/\mu$. We treat the power divergences by subtracting
the known divergences analytically from the NRQCD integrands before
performing the loop integrations.

This is enough to make the spin-dependent part well-defined.
In the spin-independent part, however, the presence of $k^2a^2$ lattice artifacts
can move the leading IR divergences to lower order, leaving an apparently
meaningless $(Ma)^2 \log(\mu a)$ overall divergence in the sum of the box
diagrams. The origin of this artifact divergence can be seen to be from the
octet Coulomb exchange; it cancels against the artifact divergence arising
from inserting the self-energy $a^2$ correction on a single Coulomb exchange
line, which also contributes to the lattice artifacts of the spin-independent
four-fermion term. Since the spin-independent part is numerically very difficult
to evaluate due to the presence of these artifact divergences, the results
for the spin-independent part are still under review.

\begin{figure}
\begin{center}
\includegraphics[width=0.5\textwidth,keepaspectratio=]{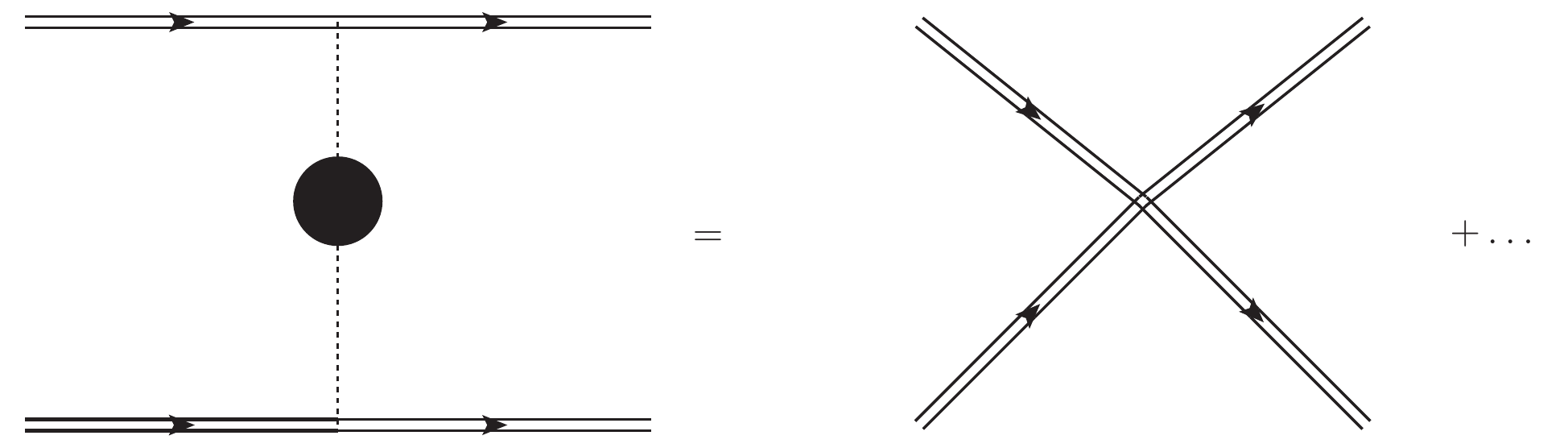}
\end{center}
\caption{A graphical depiction of the lattice artifact contribution from
         the insertion of the gluon self-energy on a single Coulomb exchange
         line.}
\end{figure}

\section{Physical Impact}

The bottomonium hyperfine splitting (HFS) receives a contribution $\sim c_4^2$ from single-gluon exchange, which is corrected by the spin-dependent four-fermion interaction $\sim (d_1-d_2)$. Empirically, the latter contribution is found to reduce the lattice-spacing dependence; a larger value of the four-fermion coefficients also partially compensates for the smaller HFS found when including spin-dependent $\calO(v^6)$ terms
\cite{Meinel:2010pv}.
The physical impact of the radiation improvement is significant: the measured 1S HFS $\sim 60$ MeV is corrected to $70$ MeV, in agreement with experiment
\cite{Dowdall:2011iy}.
In addition, the HPQCD prediction
\cite{Dowdall:2011wh}
of $35(3)(1)$ MeV for the 2S HFS including the radiative corrections agrees with the newest Belle results
\cite{Mizuk:2012pb}.

In the heavy-light case, the HFS $\sim c_4$ receives no four-fermion contribution. HPQCD finds good agreement with experiment for the $B_d$, $B_s$ HFS when including the radiative corrections to $c_4$, and is able to make a prediction for the $B_c$ HFS
\cite{Dowdall:2012ab}.

The impact of the Darwin term is much less noticeable: the bottomonium S-wave energy shift $\sim c_2$ constitutes a very small effect except on very coarse lattices.

\section{Conclusions}

We have extended the radiative improvement of $\sigma\cdot B$ term and spin-dependent four-fermion terms in the NRQCD action to the $n=4$ $v^4$ and $v^6$ actions, and have computed the radiative improvement of the NRQCD Darwin term for the $n=4$ $v^4$ action. We are currently computing the Darwin term of the $v^6$ action, as well as the spin-independent four-fermion terms.

The spin-dependent terms have a significant physical impact: in fact, the agreement of the theory with experiment depends on radiative improvement. The spin-independent terms are more subtle to compute, but have much smaller effects on heavy-quark spectra.


\footnotesize
\vskip3ex
{\bf Acknowledgements:}
We thank the DEISA Consortium,
co-funded through the EU~FP6 project RI-031513 and the FP7 project RI-222919,
for support within the DEISA Extreme Computing Initiative.
This work was supported by STFC under grants ST/G000581/1 and ST/H008861/1.
GMvH was supported in part by the DFG in the SFB~1044.
The calculations for this work were, in part, performed on the University
of Cambridge HPCs as a component of the DiRAC facility jointly funded by
STFC and the Large Facilities Capital Fund of BIS.



\begin{thebibliography}{10}

\bibitem{Lepage:1992tx}
  G.~P.~Lepage, L.~Magnea, C.~Nakhleh, U.~Magnea and K.~Hornbostel,
  Phys.\ Rev.\ D {\bf 46} (1992) 4052
  [hep-lat/9205007].

\bibitem{Hammant:2011bt}
  T.~C.~Hammant, A.~G.~Hart, G.~M.~von Hippel, R.~R.~Horgan and C.~J.~Monahan,
  Phys.\ Rev.\ Lett.\  {\bf 107} (2011) 112002
  [arXiv:1105.5309].

\bibitem{DeWitt:etc}
  B.~S.~DeWitt,
  Phys.\ Rev.\  {\bf 162} (1967) 1195;
  B.~S.~DeWitt,
  Phys.\ Rev.\  {\bf 162} (1967) 1239;
  H.~Kluberg-Stern and J.~B.~Zuber,
  Phys.\ Rev.\ D {\bf 12} (1975) 482;
  A.~Rebhan,
  Nucl.\ Phys.\ B {\bf 288} (1987) 832.

\bibitem{Hippy:etc}
  T.~C.~Hammant, R.~R.~Horgan, C.~J.~Monahan, A.~G.~Hart, E.~H.~M\"uller, A.~Gray, K.~Sivalingham and G.~M.~von Hippel,
  PoS LATTICE {\bf 2010} (2010) 043
  [arXiv:1011.2696];
  A.~Hart, G.~M.~von Hippel, R.~R.~Horgan and E.~H.~M\"uller,
  Comput.\ Phys.\ Commun.\  {\bf 180} (2009) 2698
  [arXiv:0904.0375].

\bibitem{Luscher:1995vs}
  M.~L\"uscher and P.~Weisz,
  Nucl.\ Phys.\ B {\bf 452} (1995) 213
  [hep-lat/9504006].

\bibitem{Meinel:2010pv}
  S.~Meinel,
  Phys.\ Rev.\ D {\bf 82} (2010) 114502
  [arXiv:1007.3966].

\bibitem{Dowdall:2011iy}
  R.~J.~Dowdall,
  PoS LATTICE {\bf 2011} (2011) 118
  [arXiv:1111.0449].

\bibitem{Dowdall:2011wh}
  R.~J.~Dowdall {\it et al.}  [HPQCD Collaboration],
  Phys.\ Rev.\ D {\bf 85} (2012) 054509
  [arXiv:1110.6887].

\bibitem{Mizuk:2012pb}
  R.~Mizuk {\it et al.}  [Belle Collaboration],
  arXiv:1205.6351.

\bibitem{Dowdall:2012ab}
  R.~J.~Dowdall, C.~T.~H.~Davies, T.~C.~Hammant and R.~R.~Horgan,
  arXiv:1207.5149.

\end{thebibliography}
\end{document}